\documentclass[twocolumn,english,aps,pra,showpacs]{revtex4}
\usepackage[T1]{fontenc}
\usepackage[latin9]{inputenc}
\usepackage{amsmath}
\usepackage{graphicx}
\usepackage{amssymb}
\usepackage{esint}

\makeatletter
\@ifundefined{textcolor}{}
{%
 \definecolor{BLACK}{gray}{0}
 \definecolor{WHITE}{gray}{1}
 \definecolor{RED}{rgb}{1,0,0}
 \definecolor{GREEN}{rgb}{0,1,0}
 \definecolor{BLUE}{rgb}{0,0,1}
 \definecolor{CYAN}{cmyk}{1,0,0,0}
 \definecolor{MAGENTA}{cmyk}{0,1,0,0}
 \definecolor{YELLOW}{cmyk}{0,0,1,0}
 }

\makeatother

\makeatother

\usepackage{babel}

\begin{document}

\title{Fulde-Ferrell pairing instability of a Rashba spin-orbit coupled
Fermi gas}

\author{Xia-Ji Liu$^{1}$}

\email{xiajiliu@swin.edu.au}

\affiliation{$^{1}$Centre for Atom Optics and Ultrafast Spectroscopy, Swinburne
University of Technology, Melbourne 3122, Australia}

\date{\today}
\begin{abstract}
We theoretically analyze the pairing instability of a three-dimensional
ultracold atomic Fermi gas towards a Fulde-Ferrell superfluid, in
the presence of Rashba spin-orbit coupling and in-plane Zeeman field.
We use the standard Thouless criterion for the onset of superfluidity,
with which the effect of pair fluctuations is partially taken into
account by approximately using a mean-field chemical potential at
zero temperature. This gives rise to an improved prediction of the
superfluid transition temperature beyond mean-field, particularly
in the strong-coupling unitary limit. We also investigate the pairing
instability with increasing Rashba spin-orbit coupling, along the
crossover from a Bardeen-Cooper-Schrieffer superfluid to a Bose-Einstein
condensate of Rashbons (i.e., the tightly bound state of two fermions
formed by strong Rashba spin-orbit coupling). 
\end{abstract}


\maketitle

\section{Introduction}

The recent experiment achievement of an unequal two-component gas
mixture of ultracold fermionic neutral atoms offers a unique opportunity
to solve some long-standing problems in quantum many-body physics
\cite{Giorgini2008,Bloch2008}. The key to ultracold atomic Fermi
gases is the incredible purity and precise control afforded over both
the interactions between particles and the confining environment.
This gives theorists an idealized test-bed for developing new models
that are not corrupted by complications due to unknown impurities
or disorder, which are often encountered with solid-state materials.
At present, the long-sought crossover from a Bardeen-Cooper-Schrieffer
(BCS) superfluid to a Bose-Einstein condensate (BEC) of tightly bounded
Cooper pairs has been realized experimentally \cite{Giorgini2008},
by using broad Feshbach resonances in a Fermi cloud of $^{40}$K or
$^{6}$Li atoms \cite{Regal2004,Zwierlein2004,Zwierlein2005}. The
realization of an exotic inhomogeneous superfluid state, the so-called
Fulde-Ferrell-Larkin-Ovchinnikov (FFLO) superfluid, in which the Cooper
pairs carry non-zero center-of-mass momentum \cite{Fulde1964,Larkin1964,Sheehy2006,Hu2006,Orso2007,Hu2007,Liu2007,Liu2008},
has also been indirectly demonstrated \cite{Liao2010}, by tweaking
the population imbalance of the two spin components \cite{Zwierlein2006,Partridge2006}.

Over the past year, a major advance in the field of ultracold atomic
Fermi gases is the creation of synthetic spin-orbit coupling through
the use of two counter-propagating Raman laser beams \cite{Wang2012,Cheuk2012,Fu2013}.
In solid-state systems, it is now widely known that spin-orbit coupling
is responsible for certain novel classes of materials, where topological
order plays a fundamental role \cite{Hasan2010,Qi2011}. Thus, ultracold
atomic Fermi gases appear to be a new paradigm to explore these new
types of topological materials. To date, a number of intriguing properties
of spin-orbit coupled Fermi gases have been addressed, including the
anisotropic bound state \cite{Vyasanakere2011a,Hu2011,Yu2011}, anisotropic
superfluidity \cite{Hu2011,Vyasanakere2011b,Jiang2011,Iskin2011,Seo2012},
enhanced pseudogap \cite{He2013}, and particularly topological superfluid
and Majorana fermions \cite{Gong2011,Liu2012a,Liu2012b,Wei2012,Hu2013}
- which may enable a form of quantum computing known as topological
computation \cite{Nayak2008}.

In this work, we theoretically investigate the possibility of inhomogeneous
FFLO superfluidity in spin-orbit coupled atomic Fermi gases. Our research
is motivated by the recent study by Zheng and co-workers \cite{Zheng2013}
who showed that the change of topology of Fermi surfaces due to spin-orbit
coupling and in-plane Zeeman field can provide a useful mechanics
for inhomogeneous superfluidity, in addition to the population imbalance
in two spin states. This enhanced inhomogeneous superfluidity was
first discussed by Barzykin and Gor'kov in the context of surface
superconductivity in solid-state materials such as WO$_{3}$:Na \cite{Barzykin2002}.
In contrast to solid-state superconductors, an important new ingredient
of ultracold atomic Fermi gases is strong interaction, which is necessary
in order to have an experimentally achievable superfluid transition
temperature. As a first step, in the previous studies by Zheng and
co-workers \cite{Zheng2013}, as well as by many others \cite{Wu2013,Liu2013,Shenoy2012,Dong2013FFSOC,Zhou2013,Hu2013FFSOC,Liu2013FFTS},
mean-field theory at \emph{zero} temperature has been adopted, leading
to a qualitative picture of inhomogeneous superfluidity at the BEC-BCS
crossover.

Here, we approach the FFLO problem through the analysis of pair fluctuations
and address the Fulde-Ferrell pairing instability of a normal Fermi
gas at \emph{finite} temperatures \cite{Liu2005,Liu2006}. This leads
to the so-called Thouless criterion for the onset of superfluidity,
which, in the weakly interacting limit, gives exactly the same superfluid
transition temperature as the mean-field approach. In the strongly
interacting BEC-BCS crossover, however, it provides an improved prediction
beyond mean-field, as the effect of pair fluctuations is partially
taken into account by using a modified chemical potential. We note
that the Thouless criterion is identical to finding out the two-particle
bound state in the presence of Fermi surfaces. Our study is therefore
a natural generalization of the previous analysis of pairing instability
from the two-particle perspective \cite{Shenoy2012,Dong2013}.

The rest of the paper is organized as follows. In the next section
(Sec. II), we present the model Hamiltonian for an ultracold atomic
Fermi gas with Rashba spin-orbit coupling and in-plane Zeeman field.
We describe the framework of pair-fluctuation analysis and the resulting
Thouless criterion. The chemical potential of a normal Fermi gas is
in general strongly affected by the pair fluctuations across the BEC-BCS
crossover. A quantitative evaluation of such a pair-fluctuation effect,
however, is extremely difficult, particularly in the presence of anisotropy
in the momentum space due to the combined effect of spin-orbit coupling
and in-plane Zeeman field. Therefore, in this work, we shall use an
approximate chemical potential, the mean-field chemical potential
at zero temperature. In Sec. III, we discuss in detail the Thouless
pairing instability towards a Fulde-Ferrell superfluid, by analyzing
the particle-particle vertex function of the Fermi cloud in the weakly
interacting limit and unitary limit. In Sec. IV, we consider the pairing
instability at the crossover from a BCS superfluid to a BEC of the
so-called Rashbons, i.e., tightly bound pairs formed by strong Rashba
spin-orbit coupling. Finally, Sec. IV is devoted to conclusions.

\section{Model Hamiltonian and Thouless criterion}

Let us consider a three-dimensional two-component Fermi gas of $^{6}$Li
or $^{40}$K atoms with Rashba spin-orbit coupling $\lambda(\sigma_{x}\hat{k}_{y}-\sigma_{y}\hat{k}_{x})$
and an in-plane Zeeman field along the $x$-direction $h\sigma_{x}$.
This configuration is yet to be experimentally realized \cite{Sau2011}.
Here $\hat{k}_{x}\equiv-i\partial_{x}$ and $\hat{k}_{y}\equiv-i\partial_{y}$
are the momentum operators, and $\sigma_{x}$ and $\sigma_{y}$ are
the Pauli matrices. We have denoted the strength of Rashba spin-orbit
coupling and of in-plane Zeeman field by $\lambda$ and $h$, respectively.
Near a broad Feshbach resonance, the interacting Fermi system is well-described
by a single-channel model Hamiltonian, \begin{align}
{\cal H=} & \int d{\bf x}\left\{ \psi^{\dagger}\left[\hat{\xi}_{{\bf k}}+\lambda(\hat{k}_{y}\sigma_{x}-\hat{k}_{x}\sigma_{y})+h\sigma_{x}\right]\psi\right.\nonumber \\
 & +U_{0}\psi_{\uparrow}^{\dagger}\left({\bf x}\right)\psi_{\downarrow}^{\dagger}\left({\bf x}\right)\psi_{\downarrow}\left({\bf x}\right)\psi_{\uparrow}\left({\bf x}\right),\end{align}
 where $\hat{\xi}_{\mathbf{k}}\equiv-\hbar^{2}\nabla^{2}/(2m)-\mu$
is the single-particle kinetic energy with atomic mass $m$, measured
with respect the chemical potential $\mu$, $\psi\left(\mathbf{x}\right)\equiv[\psi_{\uparrow}\left({\bf x}\right),\psi_{\downarrow}\left({\bf x}\right)]$
denotes collectively the annihilation field operators for atoms in
the spin-state $\sigma=\uparrow,\downarrow$, $U_{0}$ is the interaction
strength of the contact interaction between atoms with unlike spins.
The use of the contact interatomic interaction leads to an ultraviolet
divergence at large momentum or high energy. To overcome such a divergence,
we express the interaction strength $U_{0}$ in terms of the \textit{s}-wave
scattering length $a_{s}$, \begin{equation}
\frac{1}{U_{0}}=\frac{m}{4\pi\hbar^{2}a_{s}}-\frac{1}{V}\sum_{{\bf k}}\frac{m}{\hbar^{2}k^{2}},\end{equation}
 where $V$ is the volume of the system. Experimentally, by sweeping
an external magnetic field across the broad Feshbach resonance, the
scattering length $a_{s}$ can be tuned precisely to arbitrary values
\cite{Bloch2008}.

\subsection{Functional path-integral approach}

To solve the model Hamiltonian, we employ the functional path integral
approach and consider the partition function \begin{equation}
{\cal Z}=\int{\cal D}\left[\psi,\bar{\psi}\right]e^{-S\left[\psi\left({\bf x},\tau\right),\bar{\psi}\left({\bf x},\tau\right)\right]},\end{equation}
 where $\beta=1/(k_{B}T)$ is the inverse temperature, and $S[\psi,\bar{\psi]}\equiv\int_{0}^{\beta}d\tau[\int d{\bf x}\sum_{\sigma}\bar{\psi}_{\sigma}(\mathbf{x})\partial_{\tau}\psi_{\sigma}(\mathbf{x})+{\cal H}(\psi,\bar{\psi)}]$
is the action obtained by replacing the field operators $\psi^{\dagger}$
and $\psi$ in ${\cal H}(\psi,\psi^{\dagger})$ with the grassmann
variables $\bar{\psi}(\mathbf{x},\tau)$ and $\psi(\mathbf{x},\tau)$,
respectively. Following the standard procedure \cite{SadeMelo1993},
we introduce the pairing field $\Delta\left({\bf x},\tau\right)$
and decouple the quartic interaction term in ${\cal H}\left(\psi,\psi^{\dagger}\right)$
into a quadratic form via Hubbard-Stratonovich transformation. By
integrating out the original fermionic fields $(\psi,\psi^{\dagger})$
and expanding the pairing field around its saddle point solution $\Delta(\mathbf{x},\tau)=\Delta_{\textrm{0}}+\delta\Delta(\mathbf{x},\tau)$,
up to the level of guassian pair fluctuations \cite{SadeMelo1993},
we obtain \begin{equation}
{\cal Z}=\int{\cal D}\left[\delta\Delta,\delta\bar{\Delta}\right]e^{-S_{eff}\left[\delta\Delta\left(\mathbf{x},\tau\right),\delta\bar{\Delta}\left(\mathbf{x},\tau\right)\right]}.\end{equation}
 In accord with the expansion of the pairing field, the effective
action $S_{eff}=S_{\textrm{mf}}+\delta S$ consists of a mean-field
saddle-point part \begin{equation}
S_{\textrm{mf}}=\int_{0}^{\beta}d\tau\int d{\bf x}\frac{-\left|\Delta_{0}\right|^{2}}{U_{0}}-\frac{1}{2}\text{Tr\ensuremath{\ln\left(-{\cal G}_{0}^{-1}\right)}}+\beta\sum_{{\bf k}}\xi_{{\bf k}}\label{eq:Smf}\end{equation}
 and a gaussian-fluctuation part \begin{equation}
\delta S=\int_{0}^{\beta}d\tau\int d{\bf x}\left[-\frac{\left|\delta\Delta\left({\bf r},\tau\right)\right|^{2}}{U_{0}}+\frac{1}{4}\text{Tr}\left({\cal G}_{0}\Sigma\right)^{2}\right].\end{equation}
 Here, the trace is over all the spin, spatial, and temporal degrees
of freedom, $\xi_{{\bf k}}=\hbar^{2}\mathbf{k}^{2}/(2m)-\mu=\epsilon_{\mathbf{k}}-\mu$,
the single-particle Green function ${\cal G}_{0}$ is given by,$ $\begin{widetext}
\begin{equation}
{\cal G}_{0}^{-1}=\left[\begin{array}{cc}
-\partial_{\tau}-\hat{\xi}_{{\bf k}}-\lambda(\hat{k}_{y}\sigma_{x}-\hat{k}_{x}\sigma_{y})-h\sigma_{x} & i\Delta_{0}\hat{\sigma}_{y}\\
-i\Delta_{0}^{*}\hat{\sigma}_{y} & -\partial_{\tau}+\hat{\xi}_{{\bf k}}-\lambda(\hat{k}_{y}\sigma_{x}+\hat{k}_{x}\sigma_{y})+h\sigma_{x}\end{array}\right]\delta\left(\mathbf{x}-\mathbf{x}'\right)\delta\left(\tau-\tau'\right),\end{equation}
 \end{widetext}and the self-energy $\Sigma$ takes the form, \begin{equation}
\Sigma=\left(\begin{array}{cc}
0 & i\delta\Delta\sigma_{y}\\
-i\delta\bar{\Delta}\sigma_{y} & 0\end{array}\right).\end{equation}

In our previous study \cite{Hu2013FFSOC}, we have investigated the
possibility of Fulde-Ferrell superfluidity, based on the mean-field
saddle-point action $S_{\textrm{mf}}$ or its corresponding mean-field
thermodynamic potential $\Omega_{\textrm{mf}}=k_{B}TS_{\textrm{mf}}$.
The saddle-point solution of the pairing field $\Delta_{0}(\mathbf{x})$
has been shown to carry a non-zero center-of-mass momentum, whose
magnitude is roughly proportional to the strength of the in-plane
Zeeman field. Here, we aim to understand the instability of a \emph{normal}
Fermi gas with respect to the Fulde-Ferrell pairing, by analyzing
the gaussian-fluctuation action $\delta S$. For this purpose, in
the following we derive the particle-particle vertex function.

\subsection{Particle-particle vertex function}

In the normal phase where the pairing field vanishes, i.e., $\Delta_{0}=0$,
the inverse single-particle Green function ${\cal G}_{0}^{-1}$ is
diagonal. In the momentum space, it can be easily inverted to give
\begin{equation}
{\cal G}_{0}\left(\mathbf{k},i\omega_{m}\right)=\left[\begin{array}{cc}
g_{+}\left(\mathbf{k},i\omega_{m}\right) & 0\\
0 & g_{-}\left(\mathbf{k},i\omega_{m}\right)\end{array}\right],\end{equation}
 where $\omega_{m}\equiv(2m+1)\pi k_{B}T$ ($\nu_{n}\equiv2n\pi k_{B}T$)
is the fermionic (bosonic) Matsubara frequency, $g_{+}$ and $g_{-}$
are given by \begin{equation}
g_{+}=\frac{\left(i\omega_{m}-\xi_{{\bf k}}\right)+\left(\lambda k_{y}+h\right)\sigma_{x}-\lambda k_{x}\sigma_{y}}{\left(i\omega_{m}-\xi_{{\bf k}}\right)^{2}-\left[\left(\lambda k_{y}+h\right){}^{2}+\lambda^{2}k_{x}^{2}\right]}\end{equation}
 and \begin{equation}
g_{-}=\frac{\left(i\omega_{m}+\xi_{{\bf k}}\right)+\left(\lambda k_{y}-h\right)\sigma_{x}+\lambda k_{x}\sigma_{y}}{\left(i\omega_{m}+\xi_{{\bf k}}\right)^{2}-\left[\left(\lambda k_{y}-h\right){}^{2}+\lambda^{2}k_{x}^{2}\right]},\end{equation}
 respectively. After some algebra, we obtain the gaussian-fluctuation
part of the action as\begin{widetext} \begin{equation}
\delta S=k_{B}T\sum_{{\bf q},i\nu_{n}}\left[-\Gamma^{-1}\left({\bf q},i\nu_{n}\right)\right]\delta\Delta({\bf q},i\nu_{n})\delta\bar{\Delta}({\bf q},i\nu_{n}),\end{equation}
 where the inverse vertex function is given by \begin{equation}
\Gamma^{-1}=\frac{1}{U_{0}}+\frac{k_{B}T}{V}\sum_{{\bf k},i\omega_{m}}\left[\frac{1/2}{\left(i\omega_{m}-E_{{\bf k},+}\right)\left(i\nu_{n}-i\omega_{m}-E_{{\bf q}-{\bf k},+}\right)}+\frac{1/2}{\left(i\omega_{m}-E_{{\bf k},-}\right)\left(i\nu_{n}-i\omega_{m}-E_{{\bf q}-{\bf k},-}\right)}-A_{res}\right],\end{equation}
 with the single-particle energy \begin{eqnarray}
E_{{\bf k},\pm} & = & \xi_{{\bf k}}\pm\sqrt{\lambda^{2}k_{x}^{2}+\left(\lambda k_{y}+h\right){}^{2}},\\
E_{{\bf q}-{\bf k},\pm} & = & \xi_{{\bf q}-{\bf k}}\pm\sqrt{\lambda^{2}\left(q_{x}-k_{x}\right)^{2}+\left(\lambda q_{y}-\lambda k_{y}+h\right){}^{2}},\end{eqnarray}
 and

\begin{equation}
A_{res}\equiv\frac{\sqrt{\lambda^{2}k_{x}^{2}+\left(\lambda k_{y}+h\right){}^{2}}\sqrt{\lambda^{2}\left(q_{x}-k_{x}\right)^{2}+\left(\lambda q_{y}-\lambda k_{y}+h\right){}^{2}}+\lambda^{2}k_{x}\left(q_{x}-k_{x}\right)+\left(\lambda k_{y}+h\right)\left(\lambda q_{y}-\lambda k_{y}+h\right)}{\left(i\omega_{m}-E_{{\bf k},+}\right)\left(i\omega_{m}-E_{{\bf k},-}\right)\left(i\nu_{n}-i\omega_{m}-E_{{\bf q}-{\bf k},+}\right)\left(i\nu_{n}-i\omega_{m}-E_{{\bf q}-{\bf k},-}\right)}.\end{equation}
 By performing explicitly the summation over $i\omega_{m}$, replacing
${\bf k}$ by ${\bf q}/2+{\bf k}$ and re-arranging the terms, we
find that

\begin{eqnarray}
\Gamma^{-1} & = & \frac{m}{4\pi\hbar^{2}a_{s}}+\frac{1}{2V}\sum_{{\bf k}}\left[\frac{f\left(E_{{\bf q}/2+{\bf k},+}\right)+f\left(E_{{\bf q}/2-{\bf k},+}\right)-1}{i\nu_{n}-E_{{\bf q}/2+{\bf k},+}-E_{{\bf q}/2-{\bf k},+}}+\frac{f\left(E_{{\bf q}/2+{\bf k},-}\right)+f\left(E_{{\bf q}/2-{\bf k},-}\right)-1}{i\nu_{n}-E_{{\bf q}/2+{\bf k},-}-E_{{\bf q}/2-{\bf k},-}}-\frac{1}{\epsilon_{{\bf k}}}\right]\nonumber \\
 &  & -\frac{1}{4V}\sum_{{\bf k}}\left[1+\frac{\left(\lambda q_{x}/2\right)^{2}+\left(\lambda q_{y}/2+h\right){}^{2}-\lambda^{2}\left(k_{x}^{2}+k_{y}^{2}\right)}{\sqrt{\lambda^{2}\left(q_{x}/2+k_{x}\right)^{2}+\left(\lambda q_{y}/2+\lambda k_{y}+h\right){}^{2}}\sqrt{\lambda^{2}\left(q_{x}/2-k_{x}\right)^{2}+\left(\lambda q_{y}/2-\lambda k_{y}+h\right){}^{2}}}\right]C_{res},\end{eqnarray}
 where $f(E)\equiv1/(e^{\beta E}+1)$ is the Fermi distribution function
and \begin{eqnarray}
C_{res} & = & +\frac{\left[f\left(E_{{\bf q}/2+{\bf k},+}\right)+f\left(E_{{\bf q}/2-{\bf k},+}\right)-1\right]}{i\nu_{n}-E_{{\bf q}/2+{\bf k},+}-E_{{\bf q}/2-{\bf k},+}}+\frac{\left[f\left(E_{{\bf q}/2+{\bf k},-}\right)+f\left(E_{{\bf q}/2-{\bf k},-}\right)-1\right]}{i\nu_{n}-E_{{\bf q}/2+{\bf k},-}-E_{{\bf q}/2-{\bf k},-}}\nonumber \\
 &  & -\frac{\left[f\left(E_{{\bf q}/2+{\bf k},+}\right)+f\left(E_{{\bf q}/2-{\bf k},-}\right)-1\right]}{i\nu_{n}-E_{{\bf q}/2+{\bf k},+}-E_{{\bf q}/2-{\bf k},-}}-\frac{\left[f\left(E_{{\bf q}/2+{\bf k},-}\right)+f\left(E_{{\bf q}/2-{\bf k},+}\right)-1\right]}{i\nu_{n}-E_{{\bf q}/2+{\bf k},-}-E_{{\bf q}/2-{\bf k},+}}.\end{eqnarray}
 \end{widetext}By integrating out the quadratic pairing-fluctuation
term of $\delta S$, we obtain the contribution of the guassian pair
fluctuations to the thermodynamic potential as \begin{equation}
\delta\Omega=k_{B}T\sum_{{\bf q},i\nu_{n}}\ln\left[-\Gamma^{-1}\left({\bf q},i\nu_{n}\right)\right].\label{eq:dOmega}\end{equation}

\subsection{Thouless criterion}

Within the approximation of keeping gaussian pair fluctuations only
\cite{SadeMelo1993,Nozieres1985,Hu2006NSR,Hu2010}, the particle-particle
vertex function $\Gamma\left({\bf q},i\nu_{n}\right)$ can be physically
interpreted as the Green function of {}``Cooper pairs''. This is
already evident in Eq. (\ref{eq:dOmega}), as the thermodynamic potential
$\Omega_{B}$ of a free bosonic Green function $\mathcal{G}_{B}$
is formally given by $\Omega_{B}=k_{B}T\sum_{{\bf q},i\nu_{n}}\ln[-\mathcal{G}_{B}^{-1}(\mathbf{q},i\nu_{n})]$.
Therefore, by neglecting the interactions between Cooper pairs, which
is consistent with the approximation of gaussian pair fluctuations,
the superfluid phase transition occurs when the particle-particle
vertex function develops a pole at zero frequency $i\nu_{n}=0$. This
is the so-called Thouless criterion \cite{Liu2006,SadeMelo1993},
\begin{equation}
\textrm{max }\Gamma^{-1}\left({\bf q},i\nu_{n}=0\right)\left|_{T=T_{c}}=0\right..\label{eq:Thouless}\end{equation}
 Here, at the superfluid transition temperature $T_{c}$, the maximum
of the inverse vertex function may not occur at zero momentum $\mathbf{q}=\mathbf{0}$.
If happens, the phase coherence arises firstly among Cooper pairs
that carry a non-zero center-of-mass momentum. This is precisely the
pairing instability towards a Fulde-Ferrell superfluid.

\subsection{Approximate chemical potential}

To use the Thouless criterion, we need to know the chemical potential
$\mu$ at the superfluid transition temperature $T_{c}$, which is
to be determined by the number of particles in the Fermi cloud, consisting
of both fermions $n_{F}=-\partial\Omega_{\textrm{mf}}/\partial\mu$
and Cooper pairs $2n_{C}=-\partial\delta\Omega/\partial\mu$. In the
strongly interacting regime, the number of Cooper pairs $n_{C}$ is
significant, leading to a sizable suppression of the chemical potential.
Within the guassian pair fluctuation theory, however, such a suppression
is very difficult to determine, as the calculations of the vertex
function $\Gamma\left({\bf q},i\nu_{n}\right)$ and consequently the
thermodynamic potential $\delta\Omega$ are now greatly complicated
by the anisotropy in the momentum space arising from spin-orbit coupling
and in-plane Zeeman field. Therefore, we consider an approximate scheme
for the chemical potential, based on the following two observations:
(1) In the superfluid phase, the temperature dependence of the chemical
potential becomes weak, even in the strongly interacting regime \cite{Ku2012}.
Thus, we may set $\mu(T_{c})\simeq\mu(T=0)$; (2) At zero temperature,
the mean-field theory provides a reasonable qualitative description
of the BEC-BCS crossover \cite{Giorgini2008}. Thus, we may approximate
$\mu(T=0)\simeq\mu_{\textrm{mf}}(T=0)$. Using these two observations,
in the end we shall approximate \begin{equation}
\mu(T_{c})\simeq\mu_{\textrm{mf}}(T=0).\end{equation}

\begin{figure}[t]

\begin{centering}
\includegraphics[clip,width=0.48\textwidth]{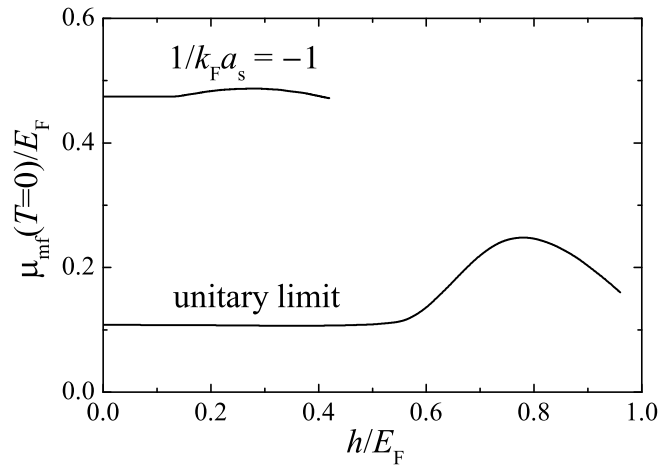} 
\par\end{centering}

\caption{(color online) Zero-temperature mean-field chemical potential of a
spin-orbit coupled Fermi gas on the BCS side ($1/k_{F}a_{s}=-1$)
and in the unitary limit ($1/k_{F}a_{s}=0$), calculated following
the approach in Ref. \cite{Hu2013FFSOC}. Here, we take a spin-orbit
coupling strength $\lambda k_{F}/E_{F}=1$, where $k_{F}$ and $E_{F}$
are the Fermi wavevector and Fermi energy, respectively.}

\label{fig1} 
\end{figure}

This approximate scheme for the chemical potential may be examined
for an ordinary BEC-BCS crossover Fermi gas without Rashba spin-orbit
coupling. In the unitary limit, where the $s$-wave scattering length
$a_{s}$ diverges, the recent accurate measurement at MIT \cite{Ku2012}
reported that $\mu(T_{c})\simeq0.42E_{F}$, in units of the Fermi
energy $E_{F}$. Although the mean-field prediction of zero temperature
chemical potential \cite{Giorgini2008}, $\mu_{\textrm{mf}}(T=0)\simeq0.59E_{F}$,
has about 40\% overestimation of $\mu(T_{c})$, it is much better
than the value commonly used in the weakly interacting limit, i.e.,
$\mu(T_{c})=E_{F}$. In Fig. \ref{fig1}, we show the zero-temperature
mean-field chemical potential of a Rashba spin-orbit coupled Fermi
gas at two dimensionless interaction parameters: $1/k_{F}a_{s}=-1$
and $1/k_{F}a_{s}=0$, to be used later in the numerical calculations.
The results are obtained by minimizing the mean-field thermodynamic
potential or action Eq. (\ref{eq:Smf}) with respect to a Fulde-Ferrell
order parameter $\Delta_{0}(\mathbf{x})=\Delta e^{iqy}$, by treating
$\Delta$ and $q$ as the independent variational parameters. For
more details, see Ref. \cite{Hu2013FFSOC}.

\section{Fulde-Ferrell pairing instability at BEC-BCS crossover}

We now determine the superfluid transition temperature of a Rashba
spin-orbit coupled Fermi gas with in-plane Zeeman field, by using
the Thouless criterion Eq. (\ref{eq:Thouless}). As discussed in the
previous section, we take the zero-temperature mean-field chemical
potential as the approximate chemical potential at $T_{c}$.

\begin{figure}[t]

\begin{centering}
\includegraphics[clip,width=0.48\textwidth]{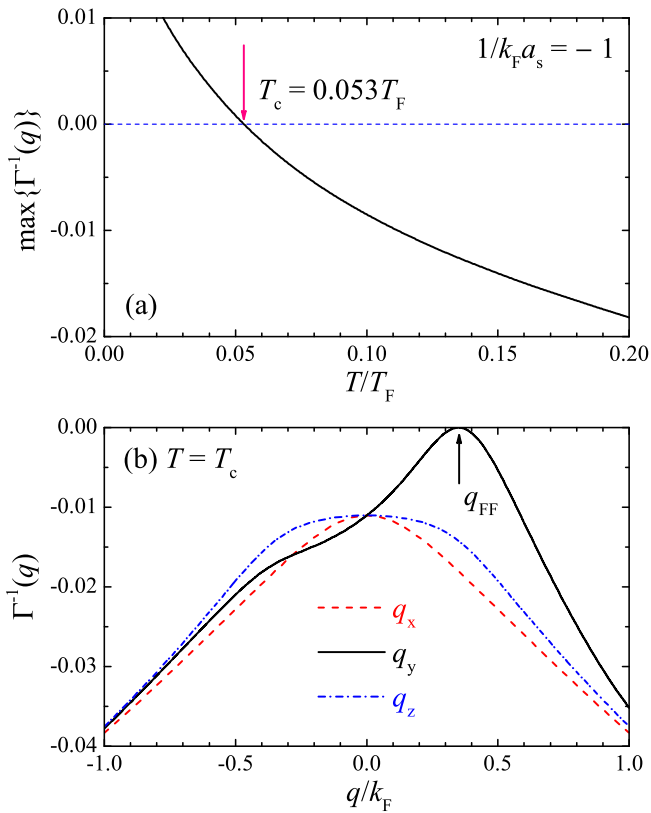} 
\par\end{centering}

\caption{(color online) (a) Maximum of the inverse particle-particle vertex
function as a function of temperature. The superfluid phase transition
occurs when the maximum touches zero, as indicated by the arrow. Here,
we consider a Rashba spin-orbit coupled Fermi gas with coupling strength
$\lambda k_{F}/E_{F}=1$ and an in-plane Zeeman field $h=0.3E_{F}$,
on the weakly interacting BCS side with an interatomic interaction
parameter $1/k_{F}a_{s}=-1$. (b) The momentum dependence of the inverse
vertex function at the superfluid transition along the $q_{x}$, $q_{y}$
and $q_{z}$ directions.}

\label{fig2} 
\end{figure}

In Fig. \ref{fig2}, we report the temperature dependence and momentum
dependence of the inverse vertex function for a weakly interacting
spin-orbit coupled Fermi gas with an in-plane Zeeman field $h=0.3E_{F}$.
The maximum of the inverse vertex function reaches zero when the temperature
decreases down to $0.053T_{F}$, indicating the onset of superfluid
transition. Remarkably, at this superfluid transition temperature,
the inverse vertex function is an anisotropic function of momentum
and its maximum occurs at a nonzero momentum $\mathbf{q}=q_{FF}\hat{e}_{y}$,
where $q_{FF}\simeq0.35k_{F}$ and $\hat{e}_{y}$ is the unit vector
along the $q_{y}$-direction. This strongly indicates that the resulting
state is an inhomogeneous Fulde-Ferrell superfluid which breaks the
spatial translation invariance. We note that the Fulde-Ferrell momentum
obtained from the Thouless criterion is consistent with the mean-field
prediction obtained in the superfluid phase at zero temperature \cite{Hu2013FFSOC},
which gives nearly the same number. This consistency is easy to understand,
as the properties of the Fermi condensate remains roughly the same
in the superfluid phase. The preference of the Fulde-Ferrell momentum
along the $q_{y}$ direction is uniquely determined by the change
of topology of the two Fermi surfaces \cite{Hu2013FFSOC}.

\begin{figure}[t]

\begin{centering}
\includegraphics[clip,width=0.48\textwidth]{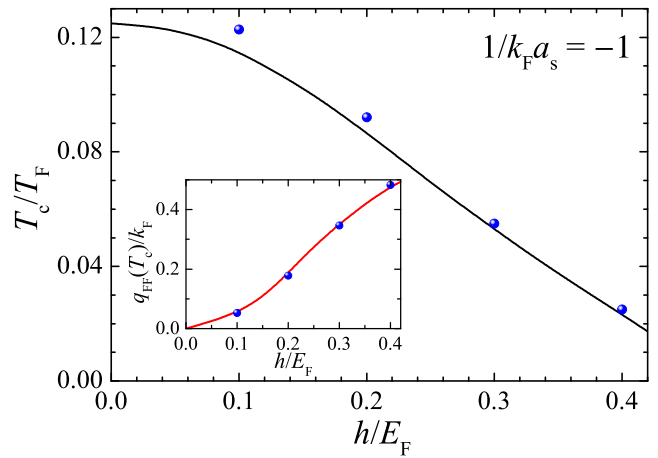} 
\par\end{centering}

\caption{(color online) Zeeman-field dependence of the superfluid transition
temperature $T_{c}$ of a spin-orbit coupled Fermi gas at the coupling
strength $\lambda k_{F}/E_{F}=1$ and at the interatomic interaction
strength $1/k_{F}a_{s}=-1$. The inset shows the Fulde-Ferrell momentum
(along the $q_{y}$ direction) at the transition as a function of
the in-plane Zeeman field. For comparison, we have also shown the
corresponding mean-field predictions by using solid circles.}

\label{fig3} 
\end{figure}

By calculating the superfluid transition temperature at different
in-plane Zeeman fields, we construct the finite temperature phase
diagram at $1/k_{F}a_{s}=-1$, as shown in Fig. \ref{fig3}. For comparison,
we also show the mean-field critical temperature at some typical Zeeman
fields by using solid circles. In this weakly interacting regime,
the chemical potential is not significantly modified by the presence
of emerging Cooper pairs. As a result, the Thouless criterion and
the mean-field calculation predict roughly the same superfluid transition
temperature, as we may anticipate. These two different approaches
also give nearly the same results for the Fulde-Ferrell momentum at
the superfluid phase transition, as illustrated by the inset of Fig.
\ref{fig3}.

\begin{figure}[t]

\begin{centering}
\includegraphics[clip,width=0.48\textwidth]{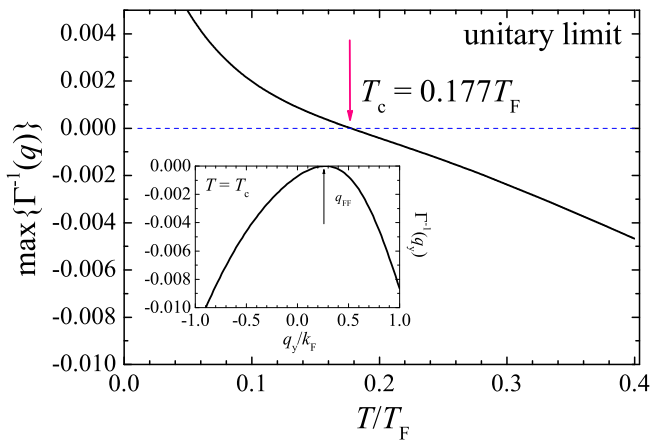} 
\par\end{centering}

\caption{(color online) The same as Fig. \ref{fig2}, except that we now consider
the strongly interacting unitary limit and we have used a large Zeeman
field $h=0.6E_{F}$. The inset shows the momentum dependence of the
inverse vertex function at the superfluid transition along the $q_{y}$
direction.}

\label{fig4} 
\end{figure}

\begin{figure}[t]

\begin{centering}
\includegraphics[clip,width=0.48\textwidth]{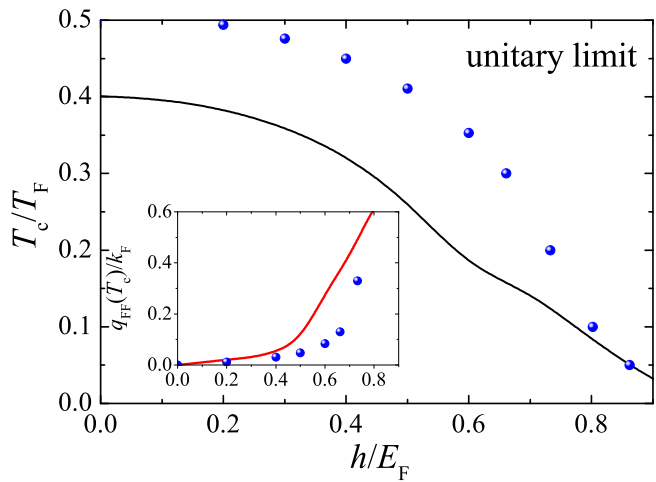} 
\par\end{centering}

\caption{(color online) Zeeman-field dependence of the superfluid transition
temperature $T_{c}$ of a spin-orbit coupled Fermi gas at the coupling
strength $\lambda k_{F}/E_{F}=1$ and in the unitary limit with a
divergent scattering length. The inset shows the Fulde-Ferrell momentum
(along the $q_{y}$ direction) at the transition as a function of
the in-plane Zeeman field. For comparison, the corresponding mean-field
predictions are shown by solid circles.}

\label{fig5} 
\end{figure}

We now turn to the strongly interacting limit. In Fig. \ref{fig4},
we show the temperature dependence and momentum dependence of the
inverse vertex function at the Zeeman field $h=0.6E_{F}$, for which
the Thouless criterion indicates that $T_{c}\simeq0.177T_{F}$. Our
calculations at different Zeeman fields lead to the determination
of the superfluid transition temperature for a spin-orbit coupled
Fermi gas in the unitary limit, as reported in Fig. \ref{fig5}. Compared
with the mean-field results (solid circles), the Thouless criterion
with the approximate chemical potential gives an improved prediction,
for in-plane Zeeman field up to $0.6E_{F}$. We could anticipate an
even smaller critical temperature when an accurate chemical potential
is used. At the typical Zeeman field $h=0.5E_{F}$, thus we estimate
that the superfluid transition may occur at about $0.2T_{F}$, a temperature
that is within the reach of the current experiment technique \cite{Ku2012}.
We note that, close to the critical Zeeman field $h_{c}\simeq E_{F}$,
above which the Fermi cloud is essentially fully polarized, the effect
of the interatomic interaction becomes much weaker. As a result, the
Thouless criterion and mean-field approach give the same results on
the superfluid transition temperature.

\begin{figure}[t]

\begin{centering}
\includegraphics[clip,width=0.48\textwidth]{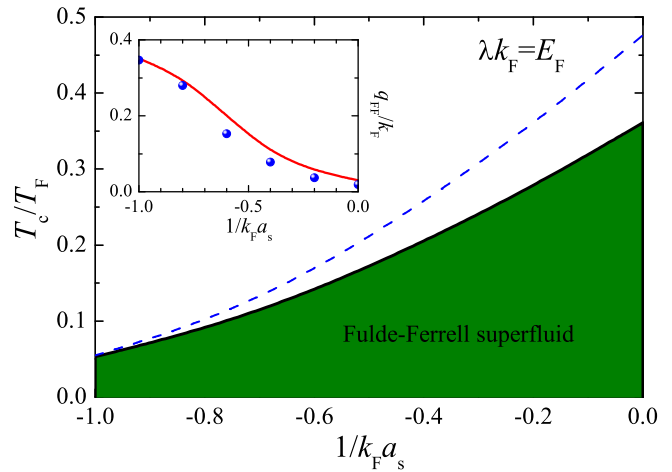} 
\par\end{centering}

\caption{(color online) Phase diagram of the BCS-BEC crossover at the in-plane
Zeeman-field $h=0.3E_{F}$ and the spin-orbit coupling strength $\lambda k_{F}/E_{F}=1$.
The critical temperatures given by Thouless criterion and mean-field
theory are shown by solid and dashed lines, respectively. The inset
shows the Fulde-Ferrell momentum at the phase transition, predicted
by the Thouless criterion (line) and the mean-field theory (solid
circles).}

\label{fig6} 
\end{figure}

In Fig. \ref{fig6}, we show the superfluid transition temperature
as a function of interaction parameter $1/k_{F}a_{s}$, at the crossover
from BCS to the unitary limit. Here we may see clearly how the prediction
by Thouless criterion starts to deviate from the mean-field result,
due to the increasing pair fluctuations.

\section{Fulde-Ferrell pairing instability of a Rashbon condensate}

We so far focus on a particular spin-orbit coupling strength $\lambda k_{F}/E_{F}=1$.
With increasing the Rashba spin-orbit coupling, it is known that a
tightly-bound pair of two fermions can form, even with a weak attractive
interatomic interaction \cite{Vyasanakere2011a,Hu2011,Yu2011}. This
new type of bound pairs, referred to as Rashbons \cite{Vyasanakere2011a,RashbonLimit},
underlies an exotic anisotropic fermionic superfluid in the absence
of Zeeman field \cite{Hu2011,Vyasanakere2011b}. Hereafter, we restrict
ourselves in the weakly interacting regime ($1/k_{F}a_{s}=-1$) and
take an in-plane Zeeman field $h=0.3E_{F}$.

\begin{figure}[t]

\begin{centering}
\includegraphics[clip,width=0.48\textwidth]{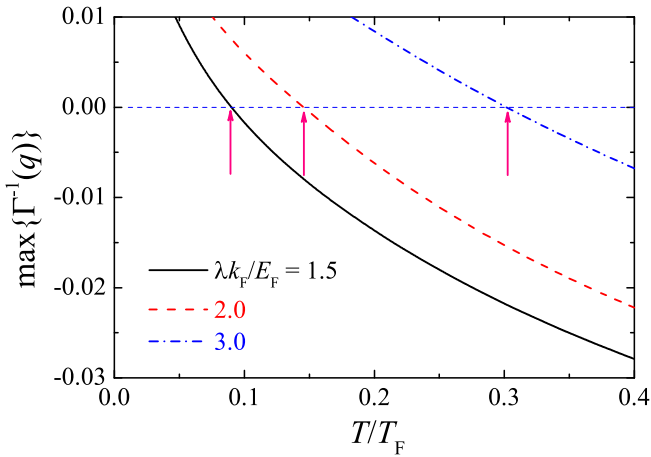} 
\par\end{centering}

\caption{(color online) Temperature dependence of the maximum of the inverse
particle-particle vertex function, at three different spin-orbit coupling
strengths, $\lambda k_{F}/E_{F}=1.5$ (solid line), $2.0$ (dashed
line) and $3.0$ (dot-dashed line), and at a weak interaction strength
$1/k_{F}a_{s}=-1$. The superfluid phase transition temperatures are
indicated by arrows. Here we take an in-plane Zeeman field $h=0.3E_{F}$.}

\label{fig7} 
\end{figure}

\begin{figure}[t]

\begin{centering}
\includegraphics[clip,width=0.48\textwidth]{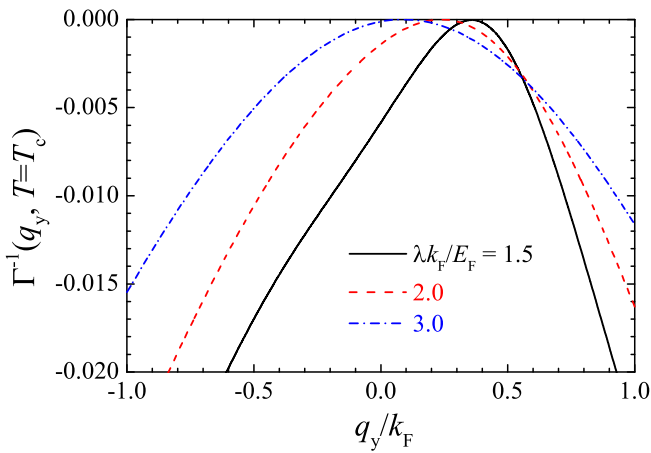} 
\par\end{centering}

\caption{(color online) The momentum dependence of the inverse vertex function
along the $q_{y}$ direction, at three different spin-orbit coupling
strengths, $\lambda k_{F}/E_{F}=1.5$ (solid line), $2.0$ (dashed
line) and $3.0$ (dot-dashed line), and at a weak interaction strength
$1/k_{F}a_{s}=-1$. We use an in-plane Zeeman field $h=0.3E_{F}$.}

\label{fig8} 
\end{figure}

In Fig. \ref{fig7}, we present the maximum of the inverse particle-particle
vertex function at the three different spin-orbit coupling strengths.
The corresponding momentum distributions along the $q_{y}$ direction
at the superfluid transition temperature are shown in Fig. \ref{fig8}.
With increasing the spin-orbit coupling, the superfluid transition
temperature increases significantly, due to the formation of Rashbons.
However, the Fulde-Ferrell momentum at the transition becomes smaller,
indicating that Fulde-Ferrell superfluidity is inherently akin to
the many-body environment.

\begin{figure}[t]

\begin{centering}
\includegraphics[clip,width=0.48\textwidth]{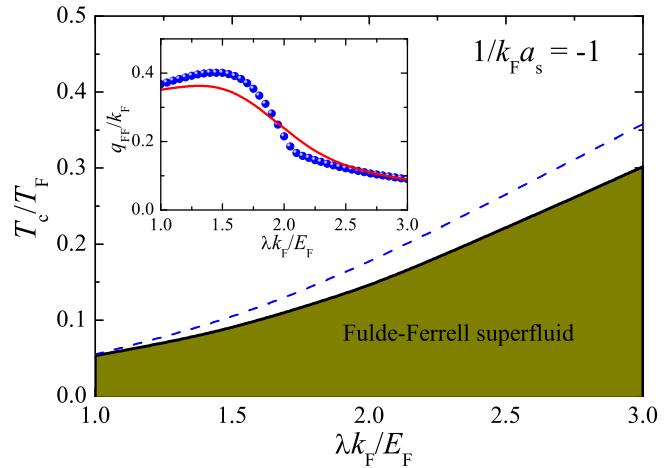} 
\par\end{centering}

\caption{(color online) Phase diagram of the crossover from a BCS superfluid
to a Rashbon BEC, at the in-plane Zeeman-field $h=0.3E_{F}$ and the
interatomic interaction strength $1/k_{F}a_{s}=-1$. The critical
temperatures given by Thouless criterion and mean-field theory are
shown by solid and dashed lines, respectively. The inset shows the
Fulde-Ferrell momentum at the phase transition, predicted by the Thouless
criterion (line) and the mean-field theory (solid circles).}

\label{fig9} 
\end{figure}

In Fig. \ref{fig9}, we present a finite-temperature phase diagram
of the crossover from a BCS superfluid to a BEC of Rashbons. As the
superfluid transition temperature increases rapidly with spin-orbit
coupling, it is already experimentally accessible at a modest coupling
strength $\lambda k_{F}/E_{F}=2$, even with a small interaction parameter
$1/k_{F}a_{s}=-1$. The Fulde-Ferrell momentum $q_{FF}$ at this coupling
strength is about $0.2k_{F}$, as shown in the inset, which might
already be large enough to be detected experimentally by, for example,
the momentum-resolved radio-frequency spectroscopy \cite{Liu2013}.
We note that, due to the use of the approximate chemical potential,
the superfluid transition temperature in the Rashbon limit is overestimated
by the Thouless criterion. It should saturate to about $0.193T_{F}$
in the limit of $\lambda\rightarrow\infty$.

\section{Conclusions}

In summary, we have investigated theoretically the Fulde-Ferrell pairing
instability in a normal, spin-orbit coupled Fermi gas with an in-plane
Zeeman field near a broad Feshbach resonance, by using the standard
Thouless criterion. In addition to complementing the existing mean-field
theoretical studies, we have predicted an improved superfluid phase
transition temperature, based on an approximate scheme for chemical
potential. We have shown that at the typical parameters, for example,
with Rashba spin-orbit coupling strength $\lambda k_{F}/E_{F}=1$,
in-plane Zeeman field $h=0.5E_{F}$ and in the unitary limit, the
Fermi cloud will become a Fulde-Ferrell superfluid at about the experimentally
attainable temperature $0.2T_{F}$. We have also presented a finite-temperature
phase diagram along the crossover from a Bardeen-Cooper-Schrieffer
superfluid to a Bose-Einstein condensate of Rashbons. 
\begin{acknowledgments}
We are grateful to Hui Hu for useful discussions. This research was
supported by the ARC Discovery Project Grant No. DP0984637 and the
NFRP-China Grant No. 2011CB921502.\end{acknowledgments}


\begin{thebibliography}{54}
\bibitem{Giorgini2008}S. Gorgini, L. P. Pitaevskii, and S. Stringari,
Rev. Mod. Phys. \textbf{80}, 1215 (2008).

\bibitem{Bloch2008}I. Bloch, J. Dalibard, and W. Zwerger, Rev. Mod.
Phys. \textbf{80}, 885 (2008).

\bibitem{Regal2004}C. A. Regal, M. Greiner, and D. S. Jin, Phys.
Rev. Lett. \textbf{92}, 040403 (2004).

\bibitem{Zwierlein2004}M. W. Zwierlein, C. A. Stan, C. H. Schunck,
S. M. F. Raupach, A. J. Kerman, and W. Ketterle, Phys. Rev. Lett.
\textbf{92}, 120403 (2004).

\bibitem{Zwierlein2005}M. W. Zwierlein, J. R. Abo-Shaeer, A. Schirotzek,
C. H. Schunck, and W. Ketterle, Nature (London) \textbf{435}, 1047
(2005).

\bibitem{Fulde1964}P. Fulde and R. A. Ferrell, Phys. Rev. \textbf{135},
A550 (1964).

\bibitem{Larkin1964}A. I. Larkin and Y. N. Ovchinnikov, Zh. Eksp.
Teor. Fiz. \textbf{47}, 1136 (1964) {[}Sov. Phys. JETP \textbf{20},
762 (1965){]}.

\bibitem{Sheehy2006}D. E. Sheehy and L. Radzihovsky, Phys. Rev. Lett.
\textbf{96}, 060401 (2006).

\bibitem{Hu2006}H. Hu and X.-J. Liu, Phys. Rev. A \textbf{73}, 051603(R)
(2006).

\bibitem{Orso2007}G. Orso, Phys. Rev. Lett. \textbf{98}, 070402 (2007).

\bibitem{Hu2007}H. Hu, X.-J. Liu, and P. D. Drummond, Phys. Rev.
Lett. \textbf{98}, 070403 (2007).

\bibitem{Liu2007}X.-J. Liu, H. Hu, and P. D. Drummond, Phys. Rev.
A \textbf{76}, 043605 (2007).

\bibitem{Liu2008}X.-J. Liu, H. Hu, and P. D. Drummond, Phys. Rev.
A \textbf{78}, 023601 (2008).

\bibitem{Liao2010}Y.-A. Liao, A. S. C. Rittner, T. Paprotta, W. Li,
G. B. Partridge, R. G. Hulet, S. K. Baur, and E. J. Mueller, Nature
\textbf{467}, 567 (2010).

\bibitem{Zwierlein2006}M. W. Zwierlein, A. Schirotzek, C. H. Schunck,
and W. Ketterle, Science \textbf{311}, 492 (2006).

\bibitem{Partridge2006}G. B. Partridge, W. Li, R. I. Kamar, Y.-A.
Liao, and R. G. Hulet, Science \textbf{311}, 503 (2006).

\bibitem{Wang2012} P. Wang, Z.-Q. Yu, Z. Fu, J. Miao, L. Huang, S.
Chai, H. Zhai, and J. Zhang, Phys. Rev. Lett. \textbf{109}, 095301
(2012).

\bibitem{Cheuk2012} L. W. Cheuk, A. T. Sommer, Z. Hadzibabic, T.
Yefsah, W. S. Bakr, and M. W. Zwierlein, Phys. Rev. Lett. \textbf{109},
095302 (2012).

\bibitem{Fu2013}Z. Fu, L. Huang, Z. Meng, P. Wang, X.-J. Liu, H.
Pu, H. Hu, and J. Zhang, Phys. Rev. A \textbf{87}, 053619 (2013).

\bibitem{Hasan2010}M. Z. Hasan and C. L. Kane, Rev. Mod. Phys. \textbf{82},
3045 (2010).

\bibitem{Qi2011}X.-L. Qi and S.-C. Zhang, Rev. Mod. Phys. \textbf{83},
1057 (2011).

\bibitem{Vyasanakere2011a}J. P. Vyasanakere and V. B. Shenoy, Phys.
Rev. B \textbf{83}, 094515 (2011).

\bibitem{Hu2011}H. Hu, L. Jiang, X.-J. Liu, and H. Pu, Phys. Rev.
Lett. \textbf{107}, 195304 (2011).

\bibitem{Yu2011}Z.-Q. Yu and H. Zhai, Phys. Rev. Lett. \textbf{107},
195305 (2011).

\bibitem{Vyasanakere2011b}J. P. Vyasanakere, S. Zhang, and V. B.
Shenoy, Phys. Rev. B \textbf{84}, 014512 (2011).

\bibitem{Jiang2011}L. Jiang, X.-J. Liu, H. Hu, and H. Pu, Phys. Rev.
A \textbf{84}, 063618 (2011).

\bibitem{Iskin2011}M. Iskin and A. L. Suba\c{s}i, Phys. Rev. Lett.
\textbf{107}, 050402 (2011).

\bibitem{Seo2012}K. Seo, L. Han, and C. A. R. Sá de Melo, Phys. Rev.
Lett. \textbf{109}, 105303 (2012).

\bibitem{He2013}L. He, X.-G. Huang, H. Hu, and X.-J. Liu, Phys. Rev.
A \textbf{87}, 053616 (2013).

\bibitem{Gong2011}M. Gong, S. Tewari, and C. Zhang, Phys. Rev. Lett.
\textbf{107}, 195303 (2011).

\bibitem{Liu2012a}X.-J. Liu, L. Jiang, H. Pu, and H. Hu, Phys. Rev.
A \textbf{85}, 021603(R) (2012).

\bibitem{Liu2012b}X.-J. Liu and H. Hu, Phys. Rev. A \textbf{85},
033622 (2012).

\bibitem{Wei2012}R. Wei and E. J. Mueller, Phys. Rev. A \textbf{86},
063604 (2012).

\bibitem{Hu2013}H. Hu, L. Jiang, H. Pu, Y. Chen, and X.-J. Liu, Phys.
Rev. Lett. \textbf{110}, 020401 (2013).

\bibitem{Nayak2008}C. Nayak, S. H. Simon, A. Stern, M. Freedman,
and S. D. Sarma, Rev. Mod. Phys. \textbf{80}, 1083 (2008).

\bibitem{Zheng2013}Z. Zheng, M. Gong, X. Zou, C. Zhang, and G.-C.
Guo, Phys. Rev. A \textbf{87}, 031602(R) (2013).

\bibitem{Barzykin2002}V. Barzykin and L. P. Gorkov, Phys. Rev. Lett.
\textbf{89}, 227002 (2002).

\bibitem{Wu2013}F. Wu, G.-C. Guo, W. Zhang, and W. Yi, Phys. Rev.
Lett. \textbf{110}, 110401 (2013).

\bibitem{Liu2013}X.-J. Liu and Hui Hu, Phys. Rev. A \textbf{87},
051608(R) (2013).

\bibitem{Shenoy2012}V. B. Shenoy, Phys. Rev. A \textbf{88}, 033609
(2013).

\bibitem{Dong2013FFSOC}L. Dong, L. Jiang, and H. Pu, New J. Phys.
\textbf{15}, 075014 (2013).

\bibitem{Zhou2013}X.-F. Zhou, G.-C. Guo, W. Zhang, and W. Yi, Phys.
Rev. A \textbf{87}, 063606 (2013).

\bibitem{Hu2013FFSOC}H. Hu and X.-J. Liu, arXiv:1304. 0387 (2013);
New J. Phys. in press.

\bibitem{Liu2013FFTS}X.-J. Liu and H. Hu, Phys. Rev. A \textbf{88},
023622 (2013).

\bibitem{Liu2005}X.-J. Liu and H. Hu, Phys. Rev. A \textbf{72}, 063613
(2005).

\bibitem{Liu2006}X.-J. Liu and H. Hu, Europhys. Lett. \textbf{75},
364 (2006).

\bibitem{Dong2013}L. Dong, L. Jiang, H. Hu, and H. Pu, Phys. Rev.
A \textbf{87}, 043616 (2013).

\bibitem{Sau2011}J. D. Sau, R. Sensarma, S. Powell, I. B. Spielman,
and S. Das, Sarma, Phys. Rev. B \textbf{83}, 140510(R) (2011).

\bibitem{SadeMelo1993}C. A. R. Sá de Melo, M. Randeria, and J. R.
Engelbrecht, Phys. Rev. Lett. \textbf{71}, 3202 (1993).

\bibitem{Nozieres1985}P. Nozières and S. Schmitt-Rink, J. Low Temp.
Phys. \textbf{59}, 195 (1985).

\bibitem{Hu2006NSR}H. Hu, X.-J. Liu, and P. D. Drummond, Europhys.
Lett. \textbf{74}, 574 (2006).

\bibitem{Hu2010}H. Hu, X.-J. Liu, and P. D. Drummond, New J. Phys.
\textbf{12}, 063038 (2010).

\bibitem{Ku2012}M. J. H. Ku, A. T. Sommer, L. W. Cheuk, and M. W.
Zwierlein, Science \textbf{335}, 563 (2012).

\bibitem{RashbonLimit}More precisely, according to the two-body picture,
a Rashbon corresponds the two-particle bound state at the limit $\hbar^{2}/(m\lambda a_{s})=0$.
Therefore, we may reach the Rashbon limit by increasing the spin-orbit
coupling strength (at a fixed interaction parameter $1/k_{F}\left|a_{s}\right|\sim1$).
\end{thebibliography}
\end{document}